\documentclass[12pt,preprint]{aastex}

\begin{document}
\title{Galaxy Size Evolution at High Redshift and Surface Brightness
Selection Effects: Constraints from the Hubble Ultra Deep Field$^{1}$}
\author{R.J. Bouwens$^{2}$, G.D. Illingworth$^{2}$,
J.P. Blakeslee$^{3}$,T.J. Broadhurst$^{4}$,M. Franx$^{5}$}
\affil{1 Based on observations made with the NASA/ESA Hubble Space
Telescope, which is operated by the Association of Universities for
Research in Astronomy, Inc., under NASA contract NAS 5-26555.}
\affil{2 Astronomy Department, University of California, Santa Cruz,
CA 95064}
\affil{3 Department of Physics and Astronomy, Johns Hopkins
University, Baltimore, MD 21218.}
\affil{4 Racah Institute of Physics, The Hebrew University, Jerusalem,
Israel 91904.}
\affil{5 Leiden Observatory, Postbus 9513, 2300 RA Leiden,
Netherlands.}

\begin{abstract}
We use the exceptional depth of the Ultra Deep Field (UDF) and
UDF-Parallel ACS fields to study the sizes of high redshift
($z\sim2-6$) galaxies and address long-standing questions about
possible biases in the cosmic star formation rate due to surface
brightness dimming.  Contrasting $B$, $V$, and $i$-dropout samples
culled from the deeper data with those obtained from the shallower
GOODS fields, we demonstrate that the shallower data are essentially
complete at bright magnitudes to $z\lesssim5.5$ and that the principal
effect of depth is to add objects at the magnitude limit.  This
indicates that high redshift galaxies are compact in size
($\sim0.1-0.3\arcsec$) and that large ($\gtrsim0.4''$,
$\gtrsim3\,$kpc) low surface brightness galaxies are rare.  A simple
comparison of the half-light radii of the HDF-N + HDF-S $U$-dropouts
with $B$, $V$, and $i$-dropouts from the UDF shows that the sizes
follow a $(1+z)^{-1.05\pm0.21}$ scaling towards high redshift.  A more
rigorous measurement compares different scalings of our $U$-dropout
sample with the mean profiles for a set of intermediate magnitude
($26.0<z_{850,AB}<27.5$) $i$-dropouts from the UDF.  The best-fit is
found with a $(1+z)^{-0.94_{-0.25} ^{+0.19}}$ size scaling (for fixed
luminosity).  This result is then verified by repeating this
experiment with different size measures, low redshift samples, and
magnitude ranges.  Very similar scalings are found for all
comparisons.  A robust measurement of size evolution is thereby
demonstrated for galaxies from $z\sim6$ to $z\sim2.5$ using data from
the UDF.
\end{abstract}
\keywords{galaxies: evolution --- galaxies: high-redshift}
\section{Introduction}

Cosmic surface brightness dimming, with its $(1+z)^4$ scaling, poses a
significant challenge to the study of high redshift galaxies (e.g.,
the bias proposed by Lanzetta et al.\ 2002).  Offsetting this is the
expectation that galaxies would be denser and therefore higher surface
brightness at high redshift (Mo, Mao, \& White 1998).  Only recently
has it become possible to explore these issues observationally
(Bouwens, Broadhurst, \& Illingworth 2003; Ferguson et al.\ 2004).  A
study of objects from the Great Observatories Origins Deep Survey
(GOODS) showed a clear decrease in size (increase in surface
brightness) from $z\sim1$ to $z\sim4$ and beyond (Ferguson et al.\
2004).  Other studies (Bouwens et al.\ 2004a; Bouwens et al.\ 2004,
hereinafter, B04) then demonstrated that the decrease extended to
$z\sim6$.  However, in extending this trend, it was necessary to make
some assumptions about the surface brightness distribution at $z\sim6$
since only the highest surface brightness objects are accessible in
GOODS at these redshifts.

In this paper, we use the exceptional depth of the Ultra Deep Field
(UDF; Beckwith et al.\ 2004) and the UDF-parallel ACS fields
(hereinafter, UDF-Ps; Bouwens et al.\ 2004a) to look at the size
(surface brightness) distribution out to $z\sim6$.  These fields reach
nearly $\sim2$ and $\sim1$ mags deeper than GOODS and for the first
time permit clean comparisons relative to lower redshift ($z\sim1-3$)
samples.  The superb depth of these fields also allows for an
important estimate of the incompleteness at high redshift in shallow,
wide area surveys like GOODS.  Throughout, we refer to the $z\sim3$
value for $L_{*}$, $M_{1700,AB}=-21.07$ (Steidel et al.\ 1999) as
$L_{*,z=3}$ and the F606W, F775W, and F850LP filters as $V_{606}$,
$i_{775}$ and $z_{850}$, respectively.  We assume
$(\Omega_{M},\Omega_{\Lambda},h)=(0.3,0.7,0.7)$ (Bennett et al.\
2003).

\section{Observations and Analysis}

To maximize our baseline for determining size changes in high redshift
galaxies ($z\sim2.5-6.0$), we adopt a $UBVi$ dropout sample set.  For
our $z\sim2.5$ $U$-dropout sample, objects are selected from the WFPC2
HDF-N and HDF-S images (B04).  For the higher redshift $z\sim3.8-6.0$
$B$, $V$, and $i$-dropout samples, objects are selected at three
different depths: one based on the relatively shallow, wide-area GOODS
fields (B04), one based on the deeper two UDF-Ps, and one based on the
UDF itself.  The selection criteria for the samples are the same ones
that were applied in B04, with magnitude limits given in Table 1 (see
B04 and Bouwens et al.\ 2004c).  These selection criteria include all
but the reddest starbursts (UV continuum slopes $\beta\lesssim 0$, or
equivalently $E(B-V)\lesssim0.45$ applied to a $10^8$ yr burst) and
some evolved galaxies (Franx et al.\ 2003) though the former objects
are expected to be rare (Adelberger \& Steidel 2000).  Contamination
from low-redshift interlopers is also expected to be small for these
samples ($\lesssim10$\%, B04).  Figure 1 shows some examples of
$i$-dropouts from the UDF.

\textit{(a) Completeness / Surface Brightness Distributions.}  Before
addressing size evolution across our sample set, it is important to
examine what effect, if any, surface brightness biases might have on
samples selected at the three depths considered here.  A convenient
way of looking at these biases is to use the size-magnitude diagram
with completeness limits overplotted.  Figure 2 shows the objects
observed from all three fields for each dropout sample using the
passband closest to rest-frame 1600 $\AA$ for the magnitude/size
measurements.  The half-light radii are calculated in circular
apertures and rely on Kron-style magnitudes (1980) (with the Kron
factor equal to 2.5) to establish the total flux.  The 50\%
completeness limits are determined from a grid of simulations over
size and magnitude.  As is clear from Figure 2, the principal effect
of the additional depth is to extend these samples to fainter
magnitudes; larger, lower surface brightness objects do not appear in
the deeper data.  This suggests that high redshift galaxies are
predominantly compact ($\sim0.1-0.3$\arcsec) and that surface
brightness biases only have a significant impact on samples close to
the magnitude limit (e.g., the GOODS $i$-dropout sample).

Binning the data in surface brightness provides us with an alternative
way of identifying biases.  Incompleteness in shallower surveys will
result in a lower surface density and a higher mean surface brightness
(reflecting the loss of the lower surface brightness population).  To
do this simultaneously with all dropout samples, we derive surface
brightness distributions over a fixed range in luminosity (0.3-1.0
$L_{*,z=3}$) (corresponding to the magnitude intervals
$24.1<V_{606,AB}<25.2$, $24.9<i_{775,AB}<26.0$, $25.4 < z_{850,AB} <
26.5$, and $26.0 < z_{850,AB} < 27.1$ for our $U$, $B$, $V$, and
$i$-dropout samples, respectively -- see the gray vertical bands in
Figure 2).  The result is plotted in Figure 3.  The surface
brightnesses for individual objects are the mean values within the
half-light radius $m_{1600,AB} + 2.5\log(2\pi r_{hl} ^2) -
2.5\log(1+z)^4$ using the mean color and redshift for each sample (see
B04) to calculate $m_{1600,AB}$.  As expected, no strong biases are
apparent for the lower redshift $B$ or $V$-dropout samples, confirming
the essential completeness of samples derived from the shallower data
sets at these magnitudes.  This situation is different however for the
$i$-dropouts as can be seen from the bias in both the mean surface
brightnesses and the surface densities: a significantly lower 17.6
mag/arcsec$^2$ in the GOODS fields vs. the 18.2 mag/arcsec$^2$ in the
UDF-Ps and 18.3 mag/arcsec$^2$ in the UDF; and a significantly lower
$0.18\pm0.02$ $i$-dropouts arcmin$^{-2}$ in the GOODS fields vs. the
$0.4\pm 0.1$ in the UDF-Ps and $0.7\pm0.2$ in the UDF, respectively.
Such a bias is not unexpected given the proximity of the GOODS
$i$-dropout sample to its completeness limit (Table 1).

\textit{(b) Size/Surface Brightness Evolution.}  Having shown that our
deeper data sets are reasonably complete at intermediate magnitudes,
we proceed to measure the size evolution out to $z\sim6$.  Before
making more rigorous estimates using a specific functional form, it is
useful just to look at how the mean size (half-light radius) varied
across our four dropout samples for objects of fixed luminosity
($0.3-1.0 L_{*,z=3}$) from Figure 2.  To minimize biases, only the UDF
is used for the $B$, $V$, and $i$-dropout samples.  Similar to the
strong trends seen at high redshift with the GOODS data (Ferguson et
al.\ 2004) where sizes decrease monotonically towards high redshift,
the present data follow a $(1+z)^{-1.05\pm0.21}$ relationship with
redshift (Figure 4).

The next step is to measure the size evolution in a more rigorous
manner giving greater emphasis to selection and measurement biases.
To do this, we use different scalings of a lower redshift sample to
model the UDF $i$-dropouts, our deepest $z\sim6$ sample.  Previously,
we used such a procedure to estimate the size evolution from the
UDF-Ps alone (Bouwens et al.\ 2004a).  Here, we take advantage of the
additional $\sim1$ mag depth available from the UDF to extend this
comparison to fainter magnitudes, $26.0<z_{850,AB}<27.5$, increasing
the size of our samples.  This magnitude range is useful since the UDF
is complete to $z_{850,AB}\sim27.5$ for objects of modest size
($<0.3\arcsec$ -- see Figure 2).  As in our previous work, we adopt
the $z\sim2.5$ HDF-N + HDF-S $U$-dropout sample as our low redshift
baseline to maximize leverage in $\Delta\log(1+z)$ and consider size
scalings of the form $(1+z)^{-m}$ where $0<m<3$, projecting the lower
redshift objects to $z\sim6$ using our well-established cloning
machinery (Bouwens et al.\ 1998a,b; Bouwens et al.\ 2003; B04), which
handles the artificial redshifting and reselection of galaxies.
Finally, before comparing against the cloned $z\sim2.5$ sample, the
UDF observations are smoothed to the $U$-dropout PSF (ACS images)
projected to $z\sim6.0$ (0.12\arcsec$\,$ FWHM).

To evaluate the validity of the different scalings, comparisons are
made using the mean radial flux profile (see B04).  This gives the
mean flux in circular annuli as a function of radius.  An illustration
of how the observations match different size scalings is provided in
Figure 5, and it is clear that the observations prefer a
$\sim(1+z)^{-1}$ size scaling of the $U$-dropouts (for fixed
luminosity).  The $(1+z)^0$ and $(1+z)^{-2}$ scalings produce profiles
which are too broad and too sharp, respectively.  Deriving the mean
and $1\sigma$ scatter expected for different scalings $m$ and
measuring the mean size from the observations (both corrected for PSF
effects), we can estimate the best-fit value for $m$, which is
$0.94_{-0.19} ^{+0.25}$.  To verify this result, the comparison was
repeated in three distinct ways: (1) making the comparison in terms of
the \textit{individual sizes} of the cloned $U$-dropouts vs. the UDF
$i$-dropouts, (2) using the mean radial flux profile of \textit{a
cloned $B$-dropout sample from the UDF} to compare with the UDF
$i$-dropouts, and (3) making the same comparison between the cloned
$U$-dropouts and UDF $i$-dropouts \textit{at fainter magnitudes}
($27<z_{850,AB}<28$).  With the possible exception of the third
comparison (where a slightly shallower scaling $m\sim0.8\pm0.2$ was
obtained), all three experiments yielded very similar scalings
($m\sim1$), suggesting that the basic result here is robust.

\section{Discussion and Summary}

In this paper, we use the exceptional depth available in the UDF and
UDF-Ps to examine the distribution of sizes and magnitudes of galaxies
at $z\sim2-6$ and contrast the results with shallow surveys like
GOODS.  We find that the principal effect of depth is to add galaxies
at faint magnitudes, not larger sizes, demonstrating that high
redshift galaxies are predominantly compact ($\sim0.1-0.3$\arcsec) and
that large ($\gtrsim0.4''$, $\gtrsim3\,$kpc) low surface brightness
objects are rare at high redshift.  The UDF therefore provides more
conclusive evidence for trends which were already apparent in the
shallower HDF + GOODS data (Bouwens et al.\ 2003; Storrie-Lombardi,
Weymann, \& Thompson 2003; Ferguson et al.\ 2004; Giavalisco et al.\
2004; B04) and HST follow-up to ground-based dropout samples
(Giavalisco, Steidel, \& Macchetto 1996) (see also discussion in
Bunker et al.\ 2004).

Contrasting galaxy sizes at the high and low redshift ends of our
sample set, we show that objects follow an approximate $(1+z)^{-1}$
relationship with redshift (for fixed luminosity).  Although
consistent, this is less steep than the $\sim(1+z)^{-1.5}$ scaling
determined at brighter ($\sim1$ mag) luminosities in our earlier
analyses (B04; Bouwens et al.\ 2004a) and hence there may be some
luminosity dependence to this scaling (and therefore evolution in the
slope of the size-magnitude relationship), see also the
$m\sim0.8\pm0.2$ scaling from the third comparison above.  Note that
the current scaling is essentially identical to the $H(z)^{-2/3}\simeq
(1+z)^{-1}$ scaling expected for systems of fixed mass (Mo et al.\
1998), pointing to a $M/L$ ratio which does not evolve much at high
redshift for $UV$-bright galaxies.  Since one can plausibly express
the UV luminosity as the gas mass divided by some star formation time
scale, one possible implication of the constant M/L ratio is one where
this time scale does not evolve much with cosmic epoch.  This is in
contrast to the steep $(1+z)^{-3/2}$ evolution in dynamical time and
suggests a scenario where feedback processes are dominant in
regulating the star formation efficiency.

Interestingly enough, a recent study (Trujillo et al.\ 2004) at lower
redshift ($0<z<3$) found that size does not evolve much with redshift
for a fixed stellar mass contrary to the Mo et al.\ (1998) scaling.
It is unknown whether this will hold true for the dynamical masses, or
how this might change at earlier times.  A resolution of these
questions will undoubtably require the measurement of these quantities
to higher redshift.

\acknowledgements

We are grateful for assistance from Dan Magee and valuable comments by
our referee Kurt Adelberger.  ACS was developed under NASA contract
NAS5-32865.  This research was supported under NASA grant
HST-GO09803.05-A and NAG5-7697.

\newpage

\begin{deluxetable}{ccccc}
\tablewidth{200pt}
\tabletypesize{\footnotesize}
\tablecaption{Dropout Samples.\label{tbl-1}}
\tablecolumns{5}
\tablehead{
\colhead{} & \colhead{Area} &
\colhead{} & \colhead{AB Mag.} & \colhead{}\\
\colhead{Sample} & \colhead{$\square\,'$} &
\colhead{\#} & \colhead{Limit\tablenotemark{a}} & 
\colhead{$L_{*}$\tablenotemark{b}}}
\startdata
$U$(HDFs) & 9 & 197 & $V \sim 26.9$ & 0.08 \\
$B$(GOODS) & 294 & 1301 & $i \sim 27.0$ & 0.14 \\
$B$(UDF-Ps) & 21 & 231 & $i \sim 28.0$ & 0.06 \\
$B$(UDF) & 13 & 248 & $i \sim 29.0$ & 0.02 \\
$V$(GOODS) & 294 & 491 & $z \sim 27.0$ & 0.23 \\
$V$(UDF-Ps) & 23 & 127 & $i \sim 28.0$ & 0.09 \\
$V$(UDF) & 13 & 160 & $i \sim 29.3$ & 0.03 \\
$i$(GOODS) & 294 & 52 & $z \sim 27.2$ & 0.33 \\
$i$(UDF-Ps) & 23 & 37 & $z \sim 28.2$ & 0.13 \\
$i$(UDF) & 13 & 85 & $z \sim 29.4$ & 0.04
\enddata
\tablenotetext{a}{Sample Selection Limit.}
\tablenotetext{b}{Limiting luminosity, using
the Steidel et al.\ (1999) value for $L_{*}$.}
\end{deluxetable}

\newpage

\begin{figure}
\epsscale{1.0}
\plotone{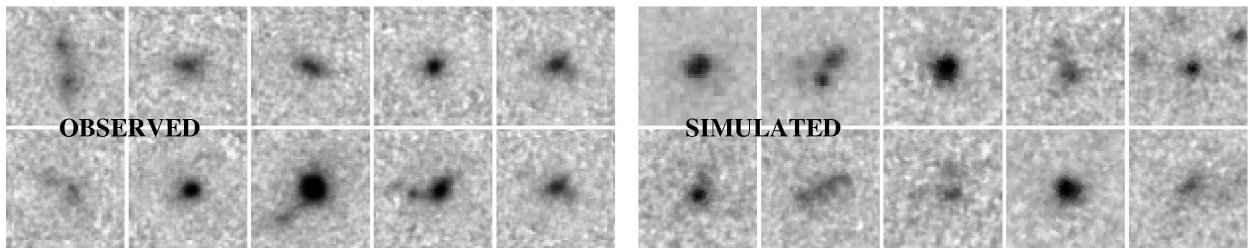}
\caption{Postage stamps ($z_{850}$ images) of the 10 brightest
($25.0<z_{850,AB}<27.2$) $i_{775}$-dropouts from the UDF compared
against a sample of HDF-N + HDF-S $U$-dropouts cloned (via
no-evolution) to $z\sim6$ and selected in a similar way.  The high S/N
of the UDF data is apparent.  For context, the object in the upper
right corner of the simulations is the familar ``quad'' from the HDF-N
(HDF4-858, Williams et al.\ 1996).}
\end{figure}

\newpage

\begin{figure}
\epsscale{1.0}
\plotone{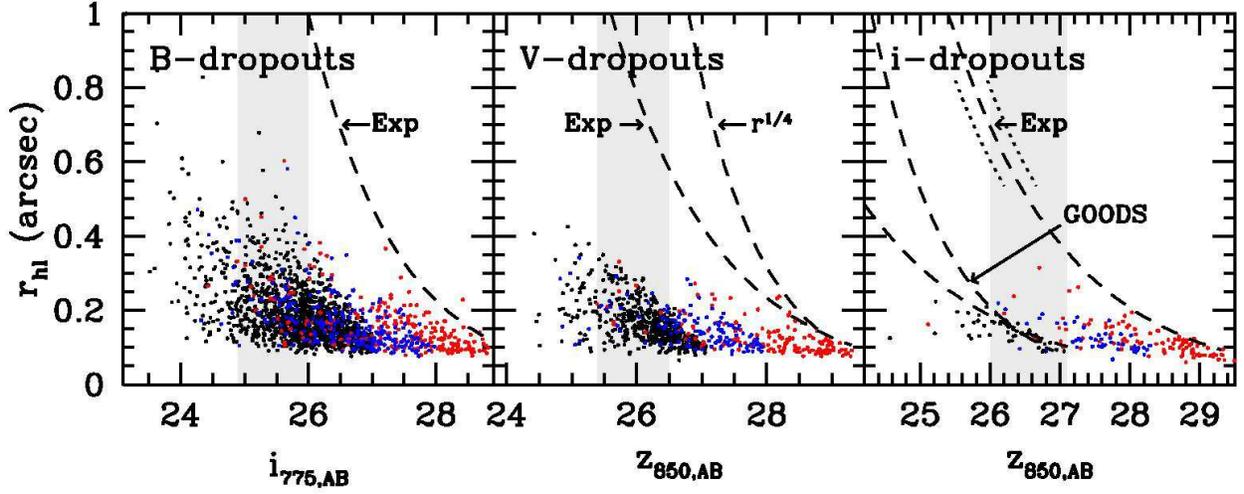}
\caption{The size-magnitude plots for the $B$, $V$, and $i$-dropout
samples extracted from GOODS (\textit{black dots}), the UDF-Ps
(\textit{blue dots}), and the UDF (\textit{red dots}).  50\%
completeness limits are shown with black dashed lines for the UDF data
(assuming an exponential surface brightness profile).  The
completeness limit for an $r^{1/4}$ surface brightness profile is also
shown.  The transition from 90\% to 10\% completeness is quite sharp
(shown in the $i$-dropout panel with the short dotted segments at 90\%
and 10\% completeness).  The magnitude range corresponding to 0.3-1.0
$L_{*,z=3}$ objects (featured in Figure 3) is indicated with the light
gray band.  Sizes are half-light radii (measured from their growth
curves).  To illustrate the severity of the selection biases on the
GOODS $i$-dropout sample, 50\% completeness limits are shown for
exponential and $r^{1/4}$ surface brightness profiles.  The principal
effect of depth is to add objects at the faint end of the surveys, not
at larger sizes, demonstrating that high redshift dropouts are
predominantly compact ($\sim0.1-0.3$\arcsec).}
\end{figure}

\begin{figure}
\epsscale{0.4}
\plotone{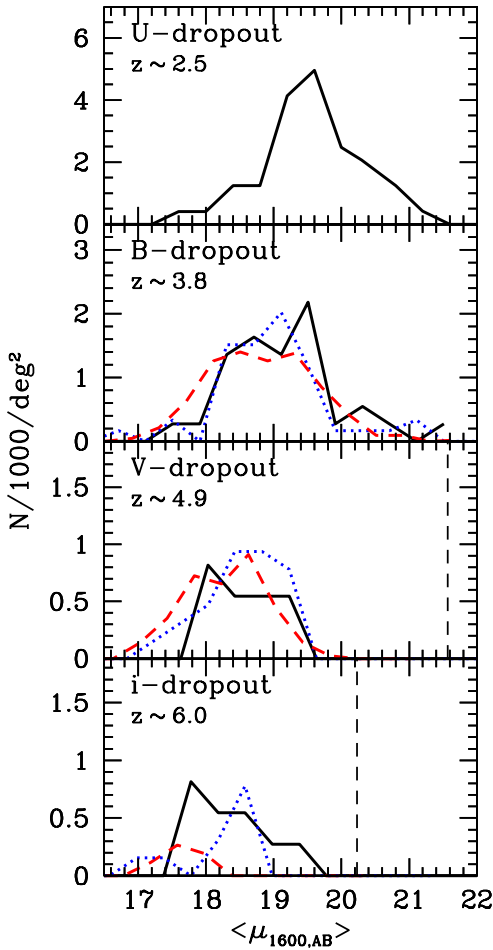}
\caption{The surface brightness distribution (corrected for surface
brightness dimming) at rest-frame $1600\AA$ for our dropout samples in
the luminosity range 0.3 $L_{*,z=3}$ to 1.0 $L_{*,z=3}$.  Shown are
the $U$-dropout sample from the HDF-N+HDF-S (\textit{top panel}) and
the $BVi$-dropout samples drawn from the GOODS fields (\textit{red
dashed lines}), the UDF-Ps (\textit{blue dotted lines}), and the UDF
(\textit{solid black lines}).  The corresponding magnitude ranges are
$24.1<i_{775,AB}<25.2$, $24.9<i_{775,AB}<26.0$,
$25.4<z_{850,AB}<26.5$, and $26.0<z_{850,AB}<27.1$ for the $U$, $B$,
$V$, and $i$-dropout samples, respectively.  The surface brightness
shown is the mean value within the half-light radius.  50\%
completeness limits for the UDF are indicated with the dashed vertical
line (calculated using an exponential surface brightness profile).
Very similar surface brightness distributions are found for $B$ and
$V$-dropouts in all three data sets, confirming that dropouts selected
from the shallower data sets are reasonably complete at the bright
magnitudes probed here.  On the other hand, for $i$-dropouts from the
GOODS fields, the surface brightness distribution (\textit{red dashed
line}) is quite biased (both in number and mean surface brightness)
relative to that obtained from the deeper data sets (UDF and UDF-Ps).
A net $\sim1.5-2.0^m$ increase in surface brightness is observed from
$z\sim2.5$ ($U$-dropouts) to $z\sim6$ ($i$-dropouts).}
\end{figure}

\begin{figure}
\epsscale{0.75}
\plotone{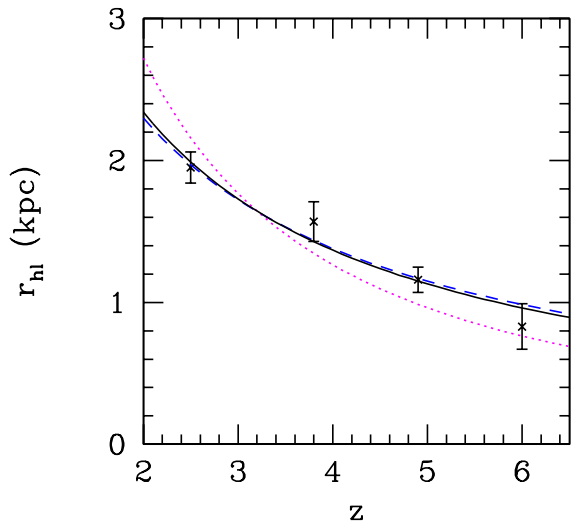}
\caption{The mean half-light radius (measured from their growth curves
and corrected for PSF effects) versus redshift for objects of fixed
luminosity ($0.3-1.0 L_{*,z=3}$).  Shown are data (\textit{crosses
with $1\sigma$ erors on the mean}) from our $z\sim2.5$ HDF-N + HDF-S
$U$-dropout sample and UDF $B$, $V$, and $i$-dropout samples plotted
at their mean redshifts $z\sim3.8$, $z\sim4.9$, and $z\sim6.0$,
respectively (see B04).  The dotted magenta line shows the
$(1+z)^{-1.5}$ scaling expected assuming a fixed circular velocity and
the dashed blue line shows the $(1+z)^{-1}$ scaling expected assuming
a fixed mass (Mo et al.\ 1998).  A least squares fit favors a
$(1+z)^{-1.05\pm0.21}$ scaling (solid black line).  \textit{This
comparison is not unbiased since objects are not selected or measured
to the same surface brightness threshold.}  The UDF is nevertheless
deep enough at these magnitudes to minimize these biases.  A more
rigorous comparison is presented in Figure 5.}
\end{figure}

\begin{figure}
\epsscale{0.75}
\plotone{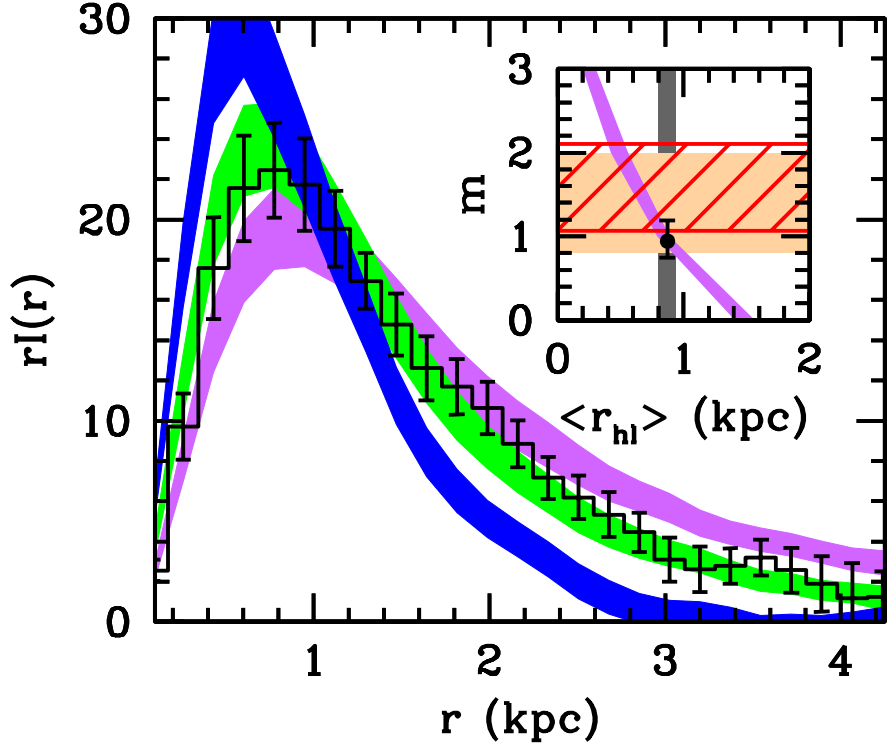}
\caption{The mean radial flux profile determined for the 15
intermediate magnitude ($26.0<z_{850,AB}<27.5$) objects from our UDF
$i$-dropout sample compared against that obtained from
similarly-selected $U$-dropouts cloned to $z\sim6$ with different size
scalings: $(1+z)^0$ (\textit{violet shading}), $(1+z)^{-1}$
(\textit{green shading}), and $(1+z)^{-2}$ (\textit{blue shading}).
The inset shows how the mean size of the projected $U$-dropouts
(\textit{shaded violet region}) vary as a function of the $(1+z)^{-m}$
size scaling exponent $m$ (a correction is made for PSF effects).
Since the mean half-light radius is $0.87\pm0.07$ kpc (\textit{shown
as a gray vertical band}), this suggests a value of $0.94_{-0.19}
^{+0.25}$ for the scaling exponent $m$.  Significantly tighter
constraints are possible on the size (surface brightness) evolution
from the UDF data than was possible in our previous study with the
UDF-Ps data (Bouwens et al.\ 2004a: \textit{red hatched region in the
inset}) and GOODS (B04: \textit{shaded orange region}) though these
probe slightly different ranges in luminosity.}
\end{figure}

\end{document}